\pdfoutput=1
\documentclass[a4paper,12pt,english]{article}

\usepackage[pdftex]{graphicx}
\usepackage{graphicx}
\usepackage{float}
\usepackage{subfigure}
\usepackage{colordvi}
\usepackage{fancybox}
\usepackage{cancel}
\usepackage{amsmath}
\usepackage{amssymb}
\usepackage{caption}
\usepackage{appendix}
\usepackage{multirow}

\usepackage{wasysym}

\usepackage{color}
\definecolor{gray}{rgb}{0.5,0.5,0.5}

\def\Tr{\mbox{Tr}\,}

\newcommand\lsim{\mathrel{\rlap{\lower4pt\hbox{\hskip1pt$\sim$}}
    \raise1pt\hbox{$<$}}}
\newcommand\gsim{\mathrel{\rlap{\lower4pt\hbox{\hskip1pt$\sim$}}
    \raise1pt\hbox{$>$}}}
\newcommand{\bet}{\begin{itemize}}
\newcommand{\eet}{\end{itemize}}
\newcommand{\beq}{\begin{equation}}
\newcommand{\eeq}{\end{equation}}
\newcommand{\bea}{\begin{eqnarray}}
\newcommand{\eea}{\end{eqnarray}}
\newcommand{\bem}{\begin{pmatrix}}
\newcommand{\eem}{\end{pmatrix}}
\newcommand{\noi}{\noindent}
\newcommand{\non}{\nonumber}

\def\lsim{\mathrel{\rlap{\lower4pt\hbox{\hskip1pt$\sim$}}
    \raise1pt\hbox{$<$}}}                
\def\gsim{\mathrel{\rlap{\lower4pt\hbox{\hskip1pt$\sim$}}
    \raise1pt\hbox{$>$}}}                

\newcommand{\drawsquare}[2]{\hbox{%
\rule{#2pt}{#1pt}\hskip-#2pt
\rule{#1pt}{#2pt}\hskip-#1pt
\rule[#1pt]{#1pt}{#2pt}}\rule[#1pt]{#2pt}{#2pt}\hskip-#2pt
\rule{#2pt}{#1pt}}

\newcommand{\Yfund}{\raisebox{-.5pt}{\drawsquare{6.5}{0.4}}}

\begin{document}

\numberwithin{equation}{section}

\begin{flushright}
\end{flushright}

\bigskip

\begin{center}

{\Large\bf    Safe SUSY  }
\vspace{1cm}

\centerline{Borut Bajc$^{a,}$\footnote{borut.bajc@ijs.si}, Nicola Andrea Dondi$^{b}$\footnote{dondi@cp3.sdu.dk} and 
Francesco Sannino$^{b,}$\footnote{sannino@cp3.sdu.dk}}

\vspace{0.5cm}
\centerline{$^{a}$ {\it\small J.\ Stefan Institute, 1000 Ljubljana, Slovenia}}
\centerline{$^{b}$ {\it\small CP$^3$-Origins \& the Danish IAS, University of Southern Denmark,  Denmark}}
 \end{center}

\bigskip

\begin{abstract}
We investigate the short distance fate of distinct classes of not asymptotically free supersymmetric gauge theories.   Examples include super QCD with two adjoint fields and generalised superpotentials, gauge theories without superpotentials and with two types of matter representation and semi-simple gauge theories such as quivers. We show that for the aforementioned theories asymptotic safety is nonperturbatively compatible with all known constraints.   
 \end{abstract}

\clearpage

\tableofcontents
 \newpage

\section{Introduction}

The  discovery of asymptotic freedom \cite{Gross:1973ju,Politzer:1973fx} has played an important role in particle physics. According to Wilson~\cite{Wilson:1971bg,Wilson:1971dh} these theories are  fundamental  since they are valid at arbitrary short and long distance scales.     Another class of fundamental theories a l\'a Wilson are the ones featuring an ultraviolet interacting fixed point, and known as asymptotically safe  theories.  The first  proof of existence of asymptotically safe gauge-Yukawa theories in four dimensions appeared in  \cite{Litim:2014uca}. These type of theories constitute now an important alternative to asymptotic freedom.  One can now imagine new extensions of the Standard Model \cite{Abel:2017ujy,Abel:2017rwl,Pelaggi:2017wzr,Mann:2017wzh,Pelaggi:2017abg
} and novel ways to achieve radiative symmetry breaking \cite{Abel:2017ujy,Abel:2017rwl}. 

In the original construction \cite{Litim:2014uca} elementary scalars and their induced Yukawa interactions play a crucial role in helping make the overall gauge-Yukawa theory safe. Quite surprisingly supersymmetric cousins of the original model, such as super QCD (SQCD) with(out) a meson and Yukawa-like superpotentials, do not support asymptotic safety  \cite{Intriligator:2015xxa}. An alleged UV fixed point, when asymptotic freedom is lost, would typically violate the $a$-theorem \cite{Cardy:1988cwa,Komargodski:2011vj,Komargodski:2011xv} inequality \cite{Intriligator:2015xxa}. It is possible to go around this constraint, as we shall see in much detail below, by considering theories with multiple fields in distinct matter representations with(out) superpotentials.  

The first such example was SQCD with two adjoints,
featuring a large enough number of quark superfields \cite{Martin:2000cr} and a superpotential. This mechanism has been recently  generalized  in \cite{Bajc:2016efj} for phenomenologically motivated SO(10) gauge theories with $3\times 16+10+210+126+\overline{126}$ \cite{Clark:1982ai,Aulakh:1982sw,Aulakh:2003kg} matter representations. The latter is dictated by the requirement that the $R$-parity \cite{Mohapatra:1986su,Font:1989ai,Martin:1992mq}  is  present at all scales 
\cite{Aulakh:1997ba,Aulakh:1998nn,Aulakh:1999cd}. These theories have all $R$-charges uniquely determined because of the presence of the superpotential and the  vanishing of the all-order NSVZ beta function \cite{Novikov:1983uc}.   One can consider vanishing superpotentials but then one has to resort to   $a$-maximization \cite{Intriligator:2003jj} to determine the $R$-charges. Explicit examples of this type appeared first in \cite{Bajc:2016efj}. 

Here we greatly enlarge these families of UV safe supersymmetric examples, and in the process we gain further insight on how to construct nonperturbatively safe supersymmetric QFTs. We also investigate quiver theories in which an interacting UV fixed point flows towards an interacting IR one. 
   
  The paper is constructed as follows: In section \ref{section2} we investigate SQCD with two adjoint fields and different superpotentials. Section \ref{section3} contains a study of SO(10) and SU(5) gauge theories with different types of vector and chiral like matter without superpotential. Quiver theories are studied in Section \ref{section4}, and we offer our conclusions in section \ref{conclusions}. 
\section{Safe SQCD with two adjoints and superpotential}
\label{section2}

In \cite{Martin:2000cr} Martin and Wells proposed a theory for which the nonperturbative existence of an interacting UV fixed point is not excluded by any known constraints. The model features the following superpotential:
\begin{equation}
W = \Tr[ \tilde{Q} X Q] + \Tr[X^3] \ , 
\end{equation}
and its field content is summarised in Table [\ref{SQCD+X+Y}]. We arrange the number of colours and flavours such that asymptotic freedom is lost and define the quantity $x = N_c/N_f$.  The vanishing of the $\beta$-function for the gauge and holomorphic coupling provides enough constraints to uniquely determine all the $R$-charges of the theory at the would be UV fixed point. Moreover, the anomalous dimensions of the  gauge singlet operators do not violate the unitarity bound. The  $\Delta a$ between the non trivial fixed point and the IR gaussian turns out to be:
\begin{equation}
\Delta a = a_{FP} - a_{FREE} =  \frac{1}{9x} (1-4x)(x-1)^2 .
\end{equation}
The non trivial UV fixed point can occur when $x < 1/4$. 

 \begin{table}[H]
\[ \begin{array}{|c|c|c c c  c|} \hline
{\rm Fields} &\left[ SU(N_c) \right] & SU_L(N_f) &SU_R(N_f) & U_V(1) & U(1)_R\\ \hline 
\hline 
W_\alpha & {\rm Adj} & 1 & 1 & 0 & 1  \\
 Q &\Yfund &\overline{\Yfund }&1& 1 &2/3  \\
\widetilde{Q}& \overline{\Yfund}&1 &  {\Yfund}& -1 & 2/3  \\ 
  X & {\rm Adj} & 1 & 1 & 0 & 2/3 \\    
   Y & {\rm Adj} & 1 & 1 & 0 & \frac{1}{3}\left( 1 + \frac{N_f}{N_c} \right) \\    
  \hline
     \end{array} 
\]
\caption{The ${\cal N}=1$ superfield content with the addition of two gauge adjoint chiral superfield $X,Y$ in the model by Martin and Wells.    }
\label{SQCD+X+Y}
\end{table} 
It can be shown that this example is part of a larger class of theories defined by the superpotential:
\begin{equation}
W  \sim \Tr[(\tilde{Q}Q)^n X^{k_1} Y^{k_2}]+ \Tr[ (\tilde{Q}Q)^m X^{l_1} Y^{l_2}] \ .
\end{equation}
For some specific choices of $n,k_1,k_2,m,l_1,l_2$. The symbol $\sim$ means that we identify all the superpotentials obtained rearranging the fields in different ways that yield the same $R$-charge constraints. The latter together with the vanishing of the NSVZ beta function gives:
\begin{equation}\begin{split}
2n R_Q + k_1 R_X + k_2 R_Y = 2 \ .\\
2m R_Q + l_1 R_X + l_2 R_Y = 2 \ ,\\
x( R_X + R_Y -1) + (R_Q -1) =  \ 0 \ . 
\end{split}\end{equation}
To avoid the emergence of free gauge invariants operators we impose:
\begin{equation}
n = 0 \wedge 2 \leq k_1 + k_2 \leq 6\quad  \vee \quad  k_1 = k_2 = 0 \wedge 1 \leq n \leq 3.
\end{equation}
We can find a total of 104 potentials providing UV fixed point satisfying all constraints. Every fixed point satisfies the constraints only in a finite $x$-interval. For example, for $x=0.46$  which is the highest possible value of $x$ allowing an UV interacting fixed point connected to the IR free one, we have seven relevant operators. These potentials read:
\begin{equation}\begin{split}
&W_1 \sim  \Tr[ X^6 ] +  \Tr[  \tilde{Q}Q X^4 ] \ ,\\
&W_2 \sim  \Tr[ X^6 ] +  \Tr[ (\tilde{Q}Q)^2 X^2 ] \ ,\\
&W_3 \sim  \Tr[ X^6 ] +  \Tr[ (\tilde{Q}Q)^3 ] \ ,\\
&W_4 \sim  \Tr[ \tilde{Q}QX^4 ] +  \Tr[  (\tilde{Q}Q)^2X^2 ] \ ,\\
&W_5 \sim  \Tr[ \tilde{Q}QX^4 ] +  \Tr[ (\tilde{Q}Q)^3] \ ,\\
&W_6 \sim  \Tr[ (\tilde{Q}Q)^2X^2] +  \Tr[  (\tilde{Q}Q)^3] \ ,\\
&W_7 \sim  \Tr[ X^5 ] +  \Tr[ (\tilde{Q}Q)^3] \ .\\
\end{split}\end{equation}
Notice that, at the fixed point, the $R$-charges are the same for the first six potentials implying that the UV value of the $a$-function is the same. Furthermore the $a$-theorem variation in between any of these UV fixed point and the trivial IR one is positive for small $x$.\\

%

\section{Safety without superpotentials: the SO(10) and SU(5) templates}
\label{section3}

We had already noticed in \cite{Bajc:2016efj} that all the known bounds for the possible existence of nonperturbative fixed points 
\bea
\label{Deltaa}
\Delta a&>&0\\ 
c&>&0\\ 
1/6&\leq&(a/c)\le1/2 \ ,
\label{noGIO}
\eea
are abided with no gauge invariant operators (GIO) with  $R<2/3$ by, for example, for an SO(10) theory featuring a very large number of generations respectively 
in the 10 and  126 representation and with vanishing superpotentials.  It is therefore timely to generalise these results.

In the following, the choice of gauge groups SO(10) or SU(5) and their representations is partially inspired by the fundamental role they play in grand unified extensions of the the Standard Model  \cite{Pati:1974yy,Georgi:1974sy,Georgi:1974yf}.  Supersymmetry is a natural playground for the unification scenario since it almost automatically predicts the correct low energy spectrum that allows for one step-unification of the 3 gauge couplings  \cite{Dimopoulos:1981yj,Ibanez:1981yh,Einhorn:1981sx,Marciano:1981un}. As discussed in \cite{Bajc:2016efj}, however,   asymptotic freedom is never respected in supersymmetric GUTs such as the ones that predict exact R-parity conservation \cite{Mohapatra:1986su,Font:1989ai,Martin:1992mq} 
at low energy \cite{Aulakh:1997ba,Aulakh:1998nn,Aulakh:1999cd}. The reason being that one needs large matter representations \cite{Aulakh:1982sw,Clark:1982ai,Babu:1992ia,Aulakh:2003kg} under SO(10), making our current investigation potentially interesting for this  line of research.

\subsection{The SO(10) template}

We start by considering susy SO(10) theories with $n_1$ generations in the representation $r_1$ and $n_2$ in the representation $r_2$ with  vanishing superpotential.

\begin{figure}[h] 
   \centering
   \includegraphics[width=3.in]{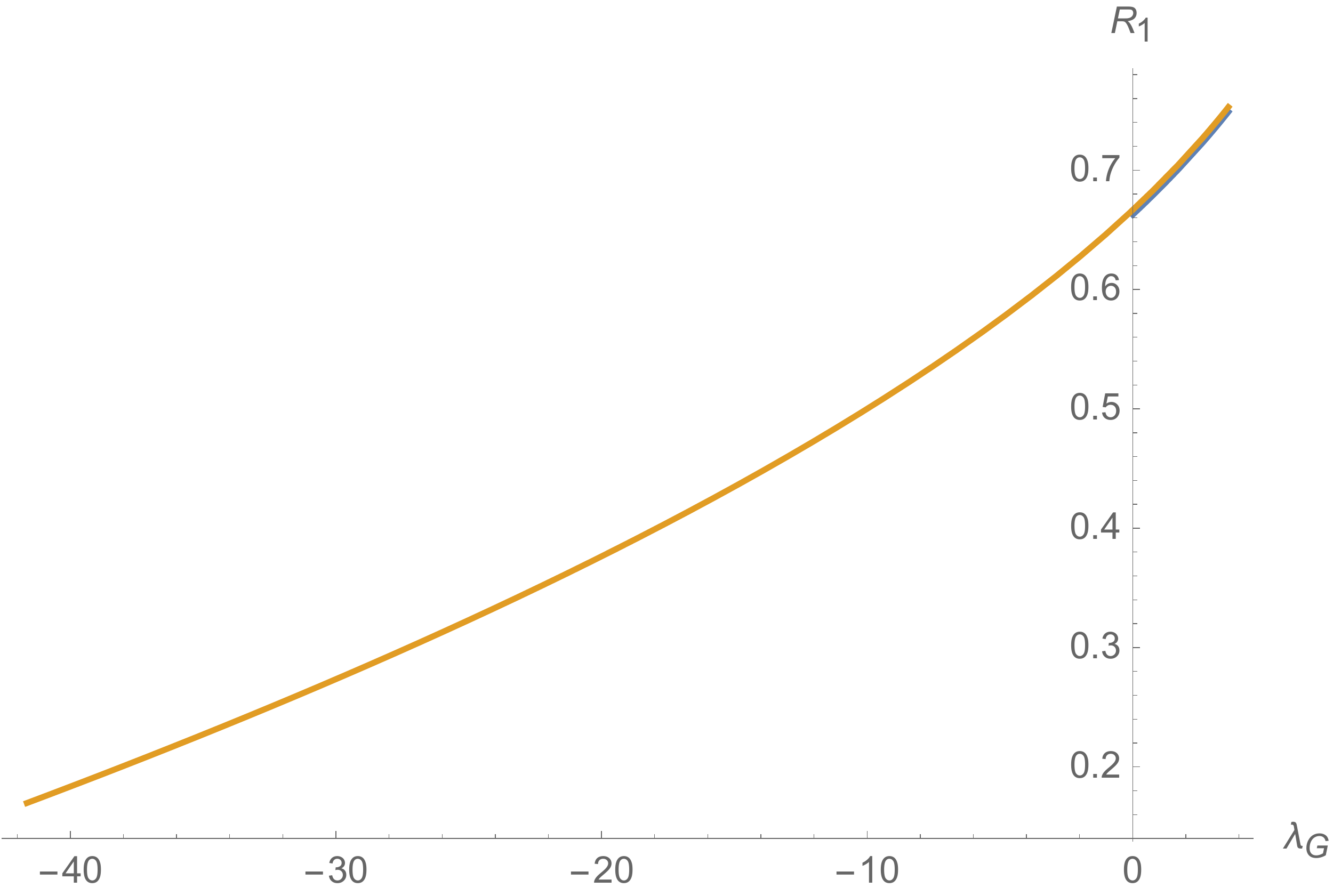} 
    \includegraphics[width=3.in]{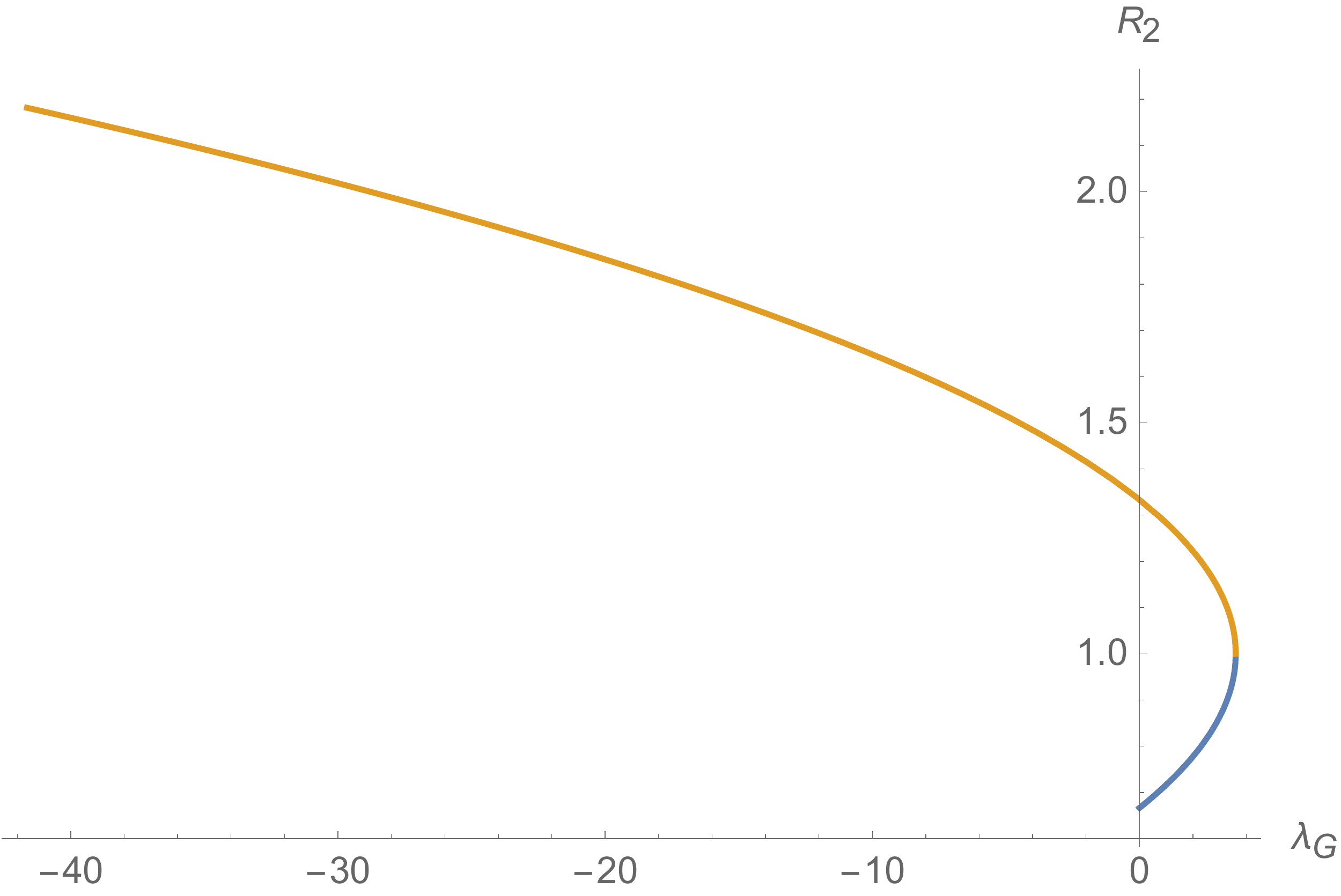} \\
   \includegraphics[width=3.in]{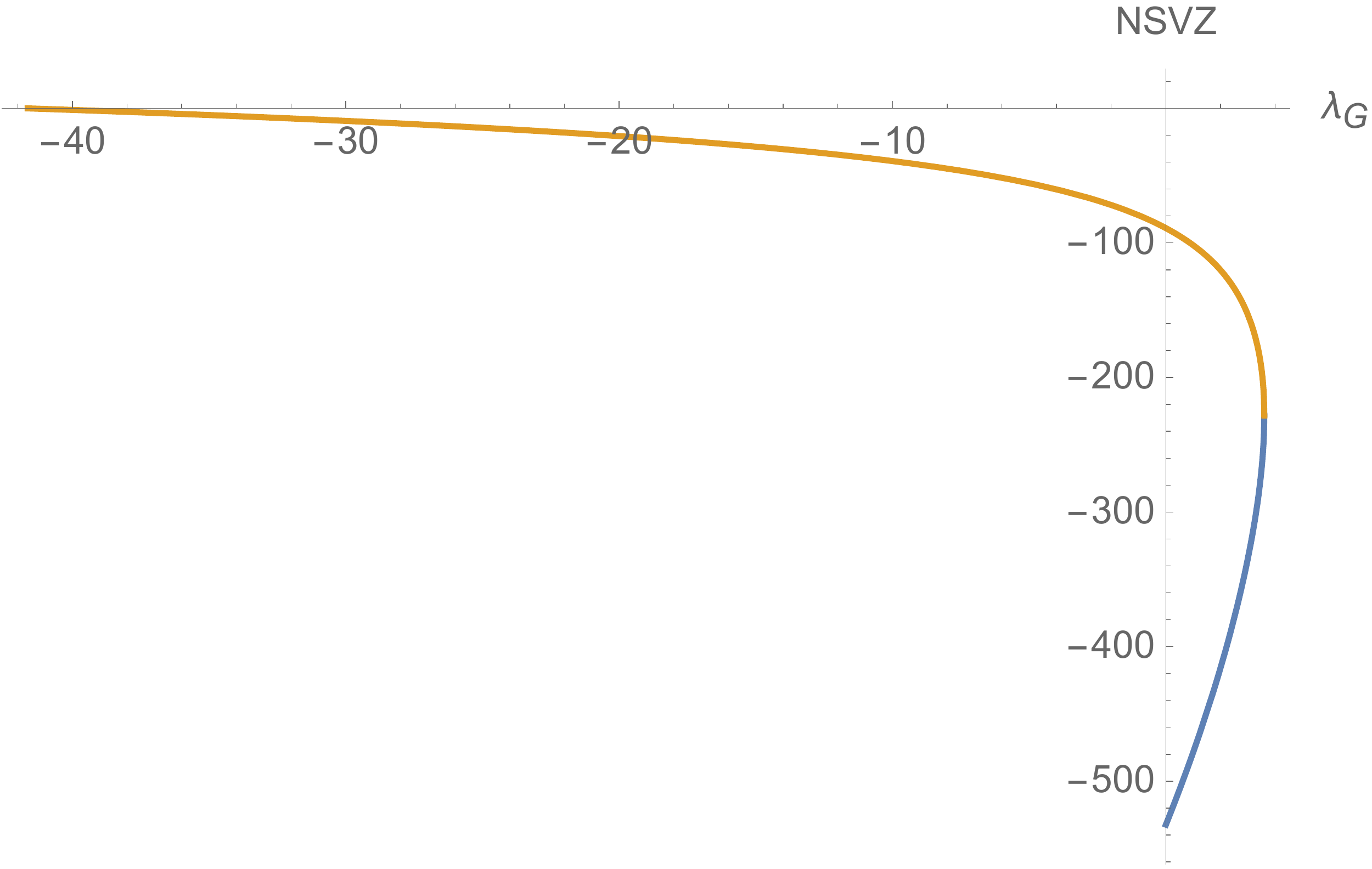} 
    \includegraphics[width=3.in]{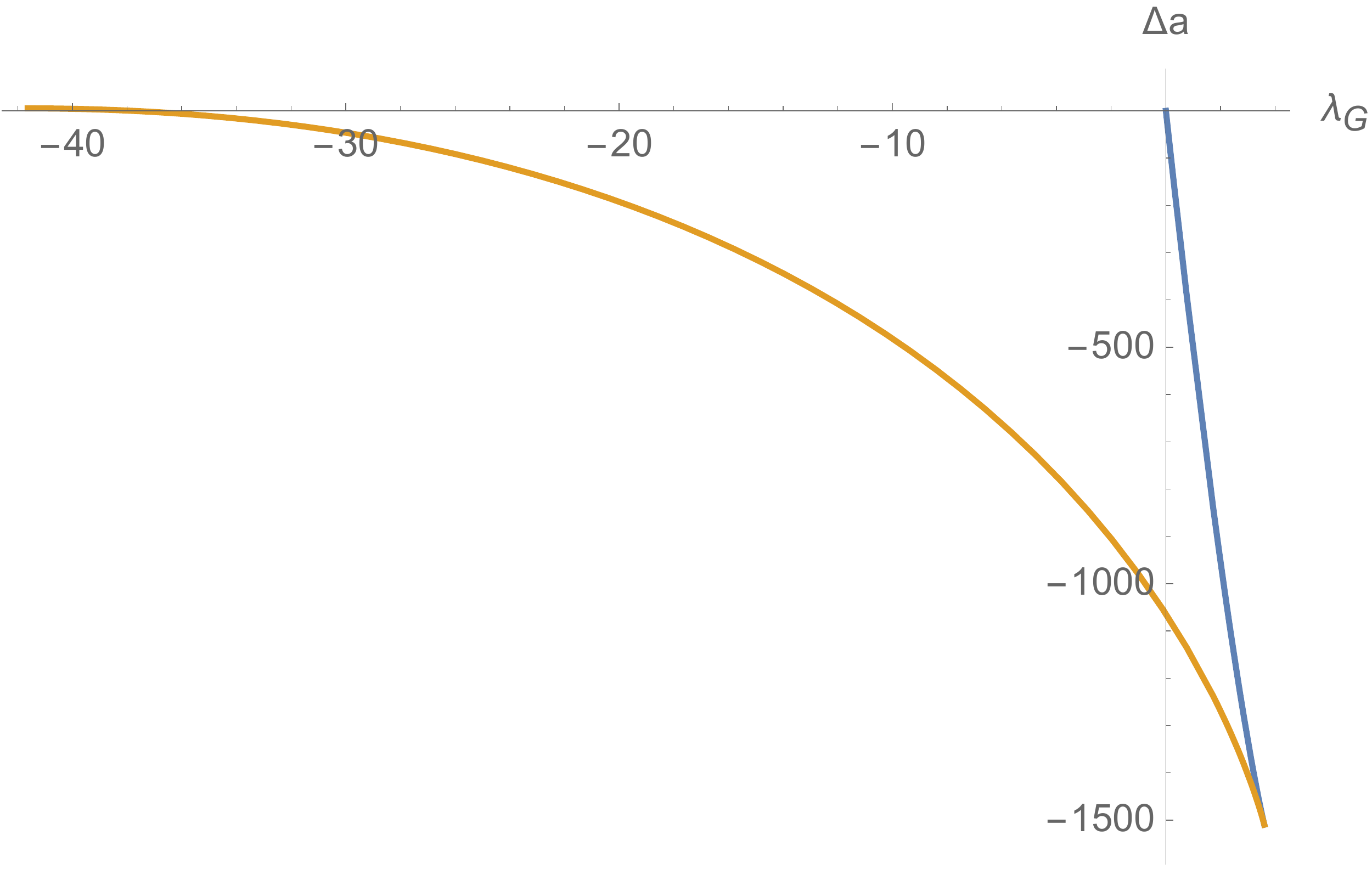} \\
   \caption{Flows of $R_1$ (upper left), $R_2$ (upper right), the $NSVZ$ function (lower left) and 
   $\Delta a$ (lower right) as functions of the Lagrange multiplier $\lambda_G$. The blue (orange) 
   curves show the direction of the flow towards increasing (decreasing) $\lambda_G$. The starting 
   points of the blue curves at $\lambda_G=0$ are at the free IR theory. }
   \label{16-126}
\end{figure}

We scan for $r_1$ and $r_2>r_1$ over the representations
\beq
10,16,45,54,120,126,144,210 \ . 
\eeq
The constraint of no GIO with $R<2/3$ is satisfied by imposing  $R>1/3$ 
for real representations and $R>1/6$ for complex representations and we discover  
 that the only solutions satisfying (\ref{Deltaa})-(\ref{noGIO}) above occur for  
\beq
\label{r1r2}
(r_1,r_2)=(10,126),\;(16,126) \ . 
\eeq
The number of generations involved is large. The reason being that to abide all the constraints one needs  at least  $n_{10} \geq 554$, while in the second case  $n_{16}\geq418$. In fact, we now argue that there is an infinite number of such solutions for integer number of generations in the 126 representation. To prove this we note that for $n_{10}\geq8490$ in the first case and for $n_{16}\geq6191$ in the 
second case there is at least one integer value of  $n_2$ for which all constraints 
(\ref{Deltaa})-(\ref{noGIO}) are satisfied. Since there is no upper bound on $n_{10}$ 
or $n_{16}$,  there is no upper bound on the number of solutions. We now turn our attention to the possibility of having a  smaller number of matter fields, but clearly still above the critical number needed to abide the constraints. We find that the most minimal among these solutions  
contains $n_1=478$ generations of 16 and $n_2=19$ generations of 126. For this example  we analyze the flow via the Lagrange multiplier technique \cite{Kutasov:2003ux} which for two type of chiral matter reduces to
\beq
a=2G+n_1r_1 a_1(R_1)+n_2 r_2 a_1(R_2)+\lambda_G\left(T_G+n_1T_1(R_1-1)+n_2T_2(R_2-1)\right)
\eeq
Extremization over $R_i$, $i=1,2$, gives
\beq
R_i(\lambda_G)=1-\frac{\epsilon_i}{3}\sqrt{1-\frac{\lambda_GT_i}{r_i}}\;\;\;,\;\;\;\epsilon_i^2=1
\eeq
In the IR ($\lambda_G=0$) the theory is free, so we are in the $\epsilon_{1,2}=+1$ branch. 
The flow goes from the IR towards positive $\lambda_G$ (that it must be positive here we know 
from perturbative calculations which are applicable for small enough $\lambda_G\sim g^2$) until it reaches 
\beq
\lambda_G^{max}\equiv Min(r_i/T_i)
\eeq
which is, in the two cases (\ref{r1r2}), always given by $126$:
\beq
\lambda_G^{max}=\frac{126}{35}=\frac{18}{5}=3.6
\eeq
At this point $\epsilon_{126}$ changes sign. $\lambda_G$ can now only decrease 
(increasing above $\lambda_G^{max}$ would lead to complex value for $R_{126}$), 
but now in the branch $\epsilon_{16}=+1$, $\epsilon_{126}=-1$. We pass through $\lambda_G=0$ 
(which is no more a free theory, because of the different branch) towards negative values of $\lambda_G$, 
all the way to the fixed point value of 
\beq
\lambda_G^*=-41.63
\eeq
for which (having $a(0)=20894/9$):
\bea
(R_{16},R_{126})(\lambda_G^*)&=&(0.1697,2.1816)\\
a(\lambda_G^*)&=&2326.5\\
\Delta a(\lambda_G^*)&=&4.955\\
c(\lambda_G^*)&=&13932.1\\
(a/c)(\lambda_G^*)-1/6&=&3.2\times10^{-4}
\eea
Notice that $4R_{16}>2/3$ but in order to avoid a free field with $R=2R_{16}<2/3$ we 
need to have only $16$ or only $\overline{16}$ but not both. This is not necessary for 
the $126$ for which the $R_{126}$ is safely large. The flows of the different quantities are shown in Fig. [\ref{16-126}].

We have therefore found an entire family of solutions that can be  asymptotically safe. Furthermore, the fact that there is a critical number of matter field value above which the asymptotically safe theory emerges within infrared gauge-matter free theories  can be viewed as the  supersymmetric analogue of the large $N_f$ solutions of non-supersymmetric safe non-abelian gauge-fermion  theories 
recently discussed in  \cite{Antipin:2017ebo}.
\subsection{The SU(5) template}

One can repeat the above analysis for  SU(5). Considering only fields up to 
representation 75, i.e. over
\beq
5,10,15,24,35,40,45,50,70,70^\prime,75
\eeq
(and their conjugates) only the following pairs can lead to consistent UV fixed point and free IR limits 
(to be on the safe side we impose here $R_i\geq1/3$ for all, real or complex representations; also, we 
assume that the $R$ charges of a field in representation $r$ is the same as the $R$ charge of an eventual 
field in the conjugate representation $\bar r$):
\beq
(r_1,r_2)=(5,35), (5,40), (5,70), (5,70^\prime), (5,75), (10,70^\prime), (24,70^\prime)
\eeq
Differently from SO(10), these SU(5) examples are not automatically anomaly free. 
To be so they must satisfy 
\beq
(n_1-n_{\bar1})A_1+(n_2-n_{\bar2})A_2=0
\eeq
where $n_a (n_{\bar a})$ is the number of generations of representation 
$r_a (\bar r_a)$, $a=1,2$, and the anomaly coefficients $A_i$ are given 
for representations $r_i$ on Table [\ref{anomalije}].
\begin{table}[]
\centering
   \begin{tabular}{|c|ccccccccccc|} 
   \hline
$r_i$  &    5 & 10 & 15 & 24 & 35 & 40 & 45 & 50 & 70 & $70^\prime$ & 75  \\
\hline
$A_i$ &     1 & 1 & 9 & 0 & -44 & -16 & -6 & -15 & 29 & -156 & 0 \\
\hline
   \end{tabular}
   \caption{\label{anomalije} The anomaly coefficients for representations smaller or equal  to 75.}
\end{table}
There are infinite number of solutions, letÕs present those with the minimal number of generations, 
taking into account only solutions with 
\beq
n_1+n_{\bar1}\leq1000\;\;\;,\;\;\;n_2+n_{\bar2}\leq5
\eeq
We summarise them in Table [\ref{resitveSU5}]. Some of them are chiral, some are vectorlike. Other solutions 
with the same number of fields are also possible. For example the second row can also have 
a vectorlike possibility with $(61+119)/2=90$ copies of $5+\bar5$ and one generation of $70+\overline{70}$. 
The opposite is not necessarily true, see the model in the last row where a chiral possibility is not present.
%
%
Notice that in order to have the precision in $\Delta a$ as specified in Table [\ref{resitveSU5}] we 
had to specify $R_{1,2}$ with higher precision, since cancellations are at work.
\begin{table}[]
\centering
   \begin{tabular}{|c|c|c|c||c|c|c|c||c|c|c|} 
   \hline
$r_1$ & $n_1$ & $n_{\bar1}$ & $R_1=R_{\bar1}$ & $r_2$ & $n_2$ & $n_{\bar2}$ & $R_2=R_{\bar2}$ & $\Delta a$ & $c_{UV}$ & $(a/c)_{UV}$ \\
\hline
\hline
5 & 15 & 147 & 0.43695 & 35 & 0 & 3 & 1.96684 & 2.15 & 1422. & 0.178 \\
5 & 61 & 119 & 0.36651 & 70 & 2 & 0 & 2.06152 & 6.38 & 1652. & 0.173 \\
5 & 9 & 165 & 0.54917 & 70' & 0 & 1 & 1.81481 & 0.75 & 1395. & 0.185 \\
5 & 90 & 90 & 0.35869 & 75 & 2 & - & 2.05436 & 0.99 & 1637. & 0.172 \\
10 & 51 & 51 & 0.43853 & 70' & 1 & 1 & 1.96316 & 13.70 & 1786. & 0.179 \\
 \hline
   \end{tabular}
   \caption{\label{resitveSU5} The candidates for UV fixed points with the minimal number of generations.}
\end{table}
%

\bigskip 
These new families of solutions show that supersymmetric gauge theories with(out) chiral matter and without superpotential can be asymptotically safe above a critical number of matter fields. Our results complement the investigation for non supersymmetric chiral gauge theories performed first in \cite{Molgaard:2016bqf}. 
Our analysis, can be straightforwardly extended to other  gauge groups  with similar matter content. One can also relax the constraints on the absence of GIO operators but this will be explored elsewhere. 

\section{Semi-simple gauge groups}
\label{section4}

Here we will analyse examples of semi-simple gauge groups starting with the quivers of  
 \cite{Intriligator:2003jj,Bertolini:2004xf}.

\subsection{The SU(N)$^4$ quiver}
The field content of the theory along with the gauge and SU(2) flavor symmetries and  charges are shown in 
Table [\ref{quiver}]. In the $N\to\infty$ limit  one recovers     the U(N)$^4$ case.
We consider a superpotential that respects all the symmetries:
\bea
\label{Wquiver}
W&=&y_1{\cal O}_1+y_2{\cal O}_2+y_3{\cal O}_3\non\\
&=&{\rm Tr}\left[ y_1\epsilon_{ab}X_{21}^aX_{14}^bX_{42}+y_2\epsilon_{ab}X_{21}^aX_{13}X_{32}^b
+y_3\epsilon_{ab}X_{21}^3X_{14}^aX_{43}X_{32}^b\right].
\eea

 \begin{table}[H]
\[ \begin{array}{|c|c c c c c|} \hline
{\rm Fields} &\left[ SU(N)_1 \right] & \left[ SU(N)_2 \right] & \left[ SU(N)_3 \right] &  \left[ SU(N)_4 \right]  & SU(2)_F \\
\hline 
\hline 
 X_{21}^a &\overline{\Yfund } & \Yfund & 1 & 1 & \Yfund  \\
X_{21}^3 &\overline{\Yfund } & \Yfund & 1 & 1 & 1  \\
X_{14}^a &\Yfund & 1 & 1 & \overline{\Yfund } & \Yfund \\
X_{43} & 1 & 1 & \overline{\Yfund } & \Yfund & 1 \\
X_{32}^a & 1 & \overline{\Yfund } &  \Yfund & 1 & \Yfund  \\
X_{13} & \Yfund & 1  & \overline{\Yfund } & 1 & 1 \\
X_{42} & 1 & \overline{\Yfund}  & 1 & \Yfund & 1 \\
\hline
     \end{array} 
\]
\caption{ The quantum numbers of the field content in the quiver example.  }
\label{quiver}
\end{table}

We study the following cases:

\begin{enumerate}

\item
\label{case1}
if $y_3=0$ there is a free field solution, with all $R_i=2/3$ and

\bea
\frac{a}{N^2}&=&\frac{92}{9}-\frac{8}{N^2}\xrightarrow{N\to\infty}10.2\\
\frac{a}{c}&=&\frac{23-18/N^2}{66-36/N^2}\xrightarrow{N\to\infty}0.348
\eea

\item
\label{case2}
if $y_1=y_2=y_3=0$ (i.e. $W=0$) we have \cite{Intriligator:2003jj}

\bea
\frac{a}{N^2}&=&\frac{2}{3}\left(3+5\sqrt{5}\right)-\frac{8}{N^2}\xrightarrow{N\to\infty}9.45\\
\frac{a}{c}&=&\frac{\frac{2}{3}\left(3+5\sqrt{5}\right)-\frac{8}{N^2}}{2\left(3+5\sqrt{5}\right)-\frac{16}{N^2}}
\xrightarrow{N\to\infty}\frac{1}{3}
\eea

The $R$-charges of the operators defined in (\ref{Wquiver}) are independent on $N$ and equal to

\beq
R({\cal O}_1)=1.87\;\;\;,\;\;\;R({\cal O}_2)=1.87\;\;\;,\;\;\;R({\cal O}_3)=2.25
\eeq

\item
\label{case3}
finally, if any of the $y_i\ne 0$ (i.e. if $W\ne 0$), we get \cite{Bertolini:2004xf}

\bea
\frac{a}{N^2}&=&\frac{32}{3}\left(-46+13\sqrt{13}\right)-\frac{8}{N^2}\xrightarrow{N\to\infty}9.30\\\
\frac{a}{c}&=&\frac{\frac{32}{3}\left(-46+13\sqrt{13}\right)-\frac{8}{N^2}}{32\left(-46+13\sqrt{13}\right)-\frac{16}{N^2}}
\xrightarrow{N\to\infty}\frac{1}{3}
\eea

The $R$-charges of the operators defined in (\ref{Wquiver}) are 

\beq
R({\cal O}_1)=2\;\;\;,\;\;\;R({\cal O}_2)=2\;\;\;,\;\;\;R({\cal O}_3)=2
\eeq

\end{enumerate}

In the limit $N\to\infty$, 
the ratio $a/c$ approaches $1/3$ for the case of point \ref{case2} and \ref{case3}, in agreement with the large $N$ expectation for  superconformal quivers  \cite{Henningson:1998gx,Benvenuti:2004dw} \footnote{The reason is \cite{Benvenuti:2004dw}, 
that in this limit  the ${\rm Tr} U(1)_R$ is proportional to the weighted sum of the NSVZ $\beta$ functions, and thus zero 
at a superconformal fixed point. Since by definition the same trace is proportional to $a-(c/3)$, the relation 
$a/c=1/3$ follows automatically for any quiver superconformal gauge theory.}. 

\noi
Requiring 

(a) $\Delta a\equiv a_{UV}-a_{IR}>0$,

(b) if $y_i\ne 0$ then $R({\cal O}_i)\geq 2$ in the IR and $R({\cal O}_i)\leq 2$ in the UV, 

\noi
we find that the only UV safe flow is for a vanishing superpotential $W=0$ in the UV (case \ref{case2}) 
and for an interacting $W\ne0$ but with $y_3=0$ in the IR (case \ref{case3}); in this scenario we have both an IR and an UV interacting 
fixed point.

\subsection{An SU(N$_1$)$\otimes$SU(N$_2$) example} 

Theories with safe trajectories for semisimple gauge groups were first analysed and discovered in  \cite{Esbensen:2015cjw}.  For these theories it is possible to achieve RG trajectories connecting UV and IR interacting fixed points.  A supersymmetric model of this type was considered in \cite{Bond:2017suy} which can be also viewed as a variant of the $SU(N)^4$ quiver in which one gauges two of the previous non-abelian flavour symmetries. We summarise in Table [\ref{LB}] the field content. 
 \begin{table}[H]
\[ \begin{array}{|c|c c c c|} \hline
{\rm Fields} &\left[ SU(N_1) \right] & \left[ SU(N_2) \right]  & SU(N_f) & SU(N_q) \\ \hline 
\hline  
{  \psi} &\Yfund  &1 & \overline{\Yfund} & 1  \\
\widetilde{ \psi}  &\overline{\Yfund } &1 & \Yfund & 1  \\
 \Psi &\Yfund &\Yfund& 1 & 1 \\
 \widetilde{ \Psi} & \overline{\Yfund } & \overline{\Yfund } &1 & 1 \\
\widetilde{ \chi } & 1 & \overline{\Yfund } &  \Yfund & 1 \\
{  \chi} & 1 & \Yfund  & \overline{\Yfund }& 1 \\
\widetilde{Q }& 1 & \overline{\Yfund}  & 1 & \Yfund \\
{Q} & 1 &\Yfund  &1 & \overline{\Yfund}\\
\hline
     \end{array} 
\]
\caption{Field content of the model introduced in  \cite{Bond:2017suy}   }
\label{LB}
\end{table} 
 
\begin{figure}[h] 
   \centering
   \includegraphics[width=3.in]{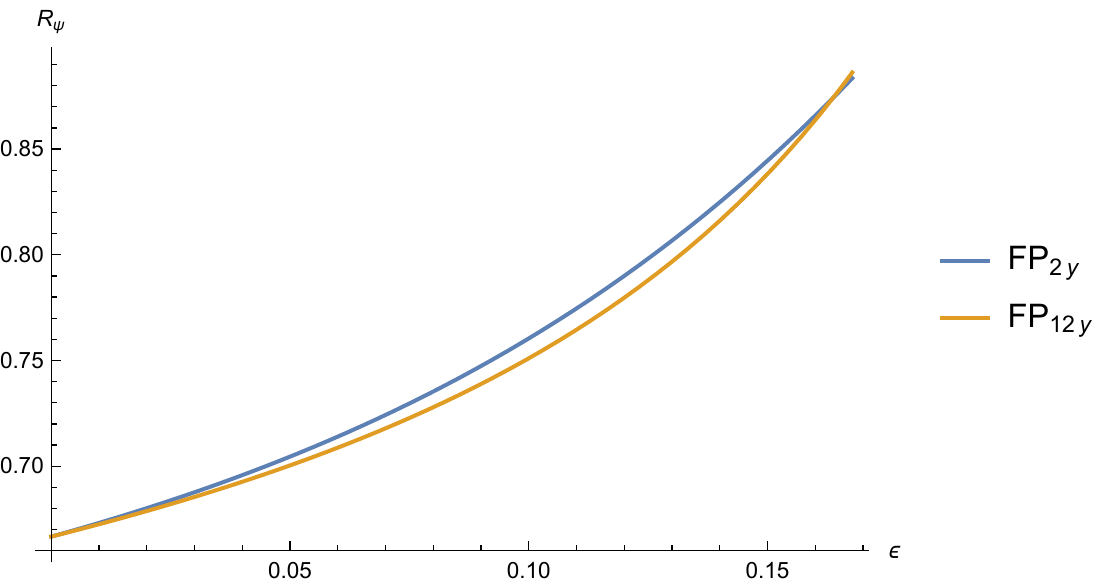} 
    \includegraphics[width=3.in]{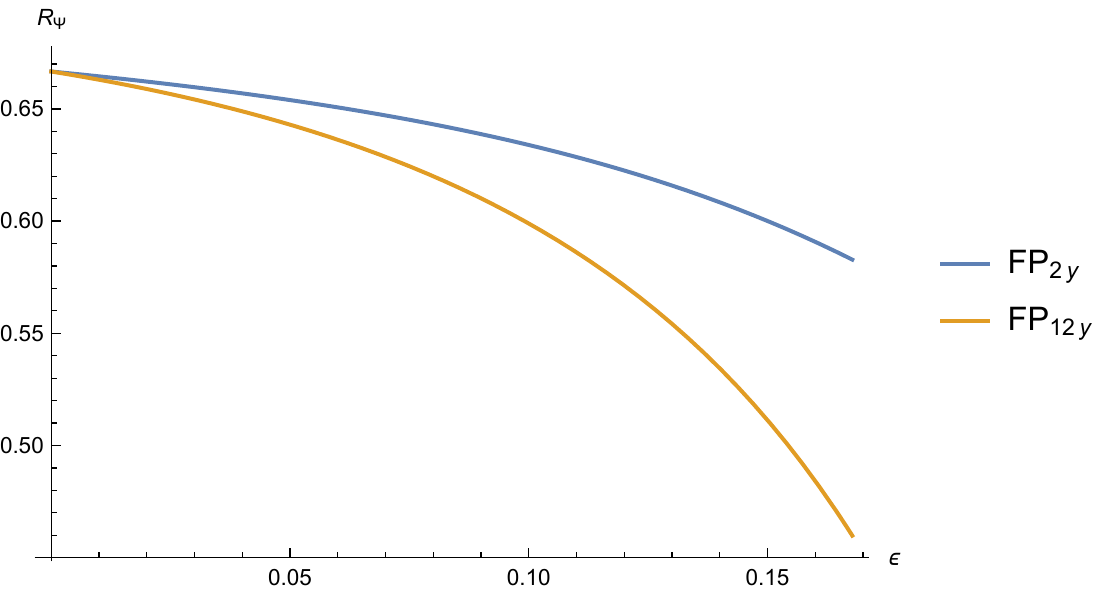} \\
   \includegraphics[width=3.in]{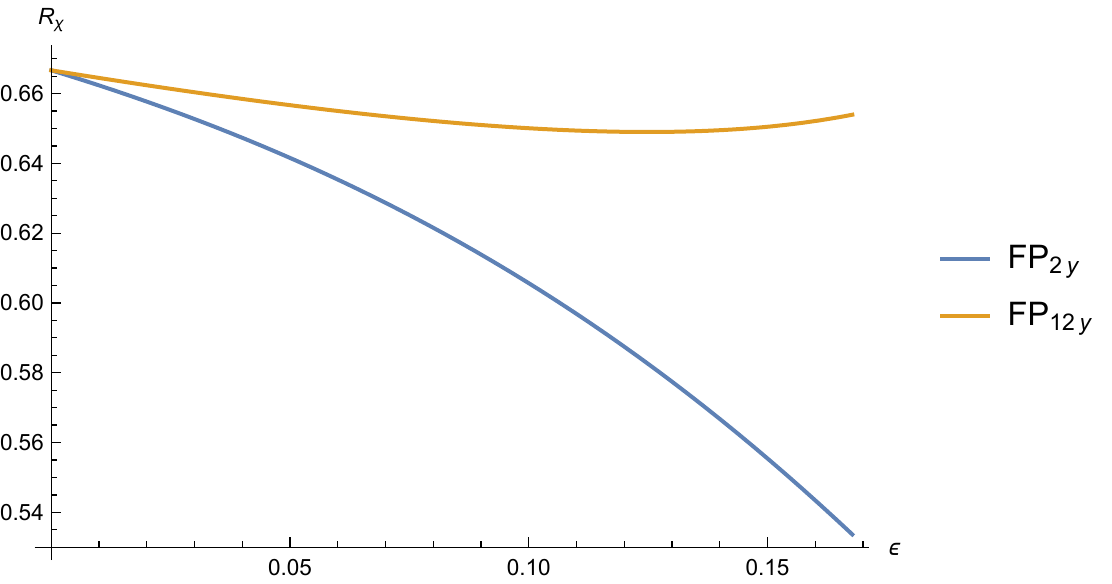} 
    \includegraphics[width=3.in]{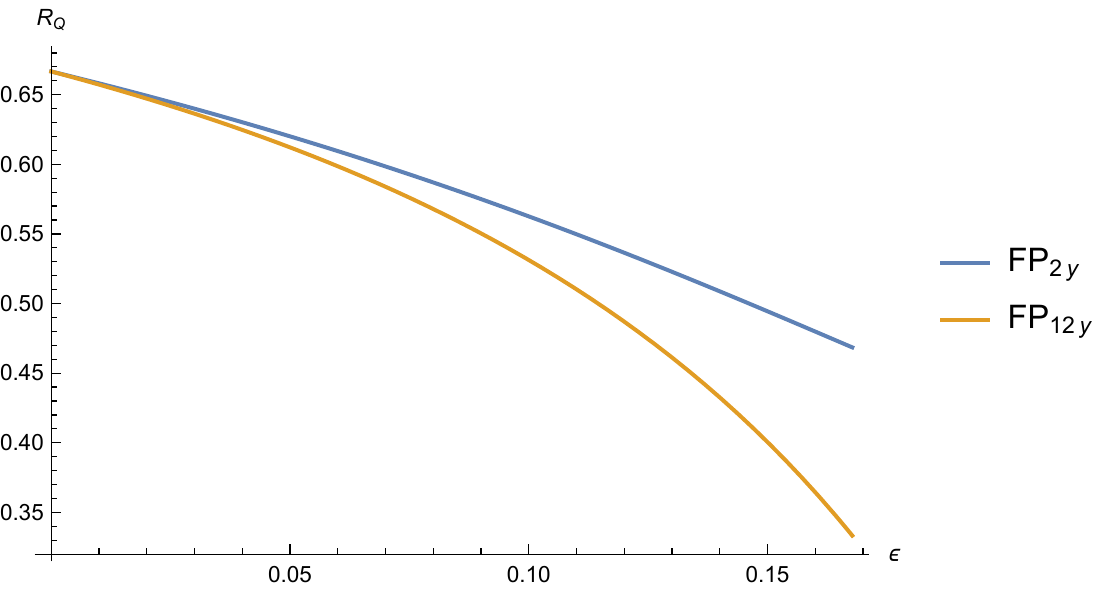} \\
   \caption{Here we draw the $R$-charges as function of the parameter $\epsilon$ for fixed 
   $P_1=3/2$ and $P_2=-5$.  }
   \label{BLR}
\end{figure}

  The model features in addition a Yukawa-type superpotential of the form:
 \beq
 W = y  \left(  \Tr[ \psi \widetilde{\Psi} \chi  ] +    \Tr[ \widetilde{\psi} {\Psi}  \widetilde{ \chi }   ] \right) \ .
 \eeq
We will consider the model in the Veneziano limit keeping the following ratios fixed:
 \begin{equation}
 x_1 = \frac{N_1}{N_f}  \ , \quad x_2 = \frac{N_2}{N_f} \ , \quad x_q = \frac{N_q}{N_f} \ .
 \end{equation}
 The $\beta$ function for the gauge and superpotential couplings are:
 \begin{equation}\begin{split}
& \beta_y = \frac{3}{2}y(R_{\psi} + R_{\Psi} + R_{\chi} - 2)\\
& \beta_1 = - \frac{3 g_1^3}{16\pi^2}f(g_1^2)[ x_1 + (R_{\psi} -1) - x_2 (R_{\Psi}-1)] \\
& \beta_2 = -\frac{3 g_2^3}{16\pi^2}f(g_2^2)[x_2 - (R_{\chi}-1)+x_1 (R_{\Psi}-1)+ x_q (R_Q -1)]\\
\end{split} \end{equation}
where $f(g^2) \sim 1 + O(g^2)$ is a scheme dependent function of the couplings. The properly normalized $a$-function reads:
 \begin{equation}\begin{split}
& a/N_f^2 = 2(x_1^2 + x_2^2) + 2 x_1 a_1(R_{\psi}) + 2 x_2 a_1 (R_{\chi}) + 2x_1 x_2 a_1 (R_{\Psi}) + 2x_q x_2 a_1 (R_Q).
\end{split} \end{equation}
We can now find the nonperturbative fixed points of the theory by setting to zero the beta functions together with a-maximisation. We also allow for partially interacting fixed points, following \cite{Esbensen:2015cjw}, meaning that some of the beta functions vanish trivially at the origin of their respective couplings. To compare our nonperturbative results with the perturbative ones given in \cite{Bond:2017suy} we introduce the further quantities:
 \begin{equation}
 P_1= \frac{x_2}{x_1}  \ , \quad P_2 = \frac{x_2}{x_1}\frac{x_q +x_1-3 x_2 +1}{x_2-3x_1 + 1} \ , \quad \epsilon = \frac{x_2 - 3x_1 +1}{x_1} \ ,
 \label{para}
 \end{equation}
and assume  $P_1=3/2$ and $P_2=-5$ while, differently from  \cite{Bond:2017suy}, our $\epsilon$ can take any positive value in the range $]0,0.16776]$ for which no free GIO can emerge. We find seven distinct potential fixed points including the fully non-interacting one in all couplings that pass all the known nonperturbative tests. Of these fixed points three are the physical ones that go over the perturbative analysis. Ordering in the descending value assumed by the central charge  $a$ these are the gaussian fixed point at the origin $G$ of all couplings, the interacting (FP$_{2y}$) in all couplings except $\alpha_1$  and the fully interacting one  (FP$_{12y}$). We report in Fig.~\ref{BLR} the nonperturbative $R$ charges for the (semi)interacting fixed points as functions of $\epsilon$. 
   
\begin{figure}[H] 
  \centering
    \includegraphics[width=0.47\textwidth]{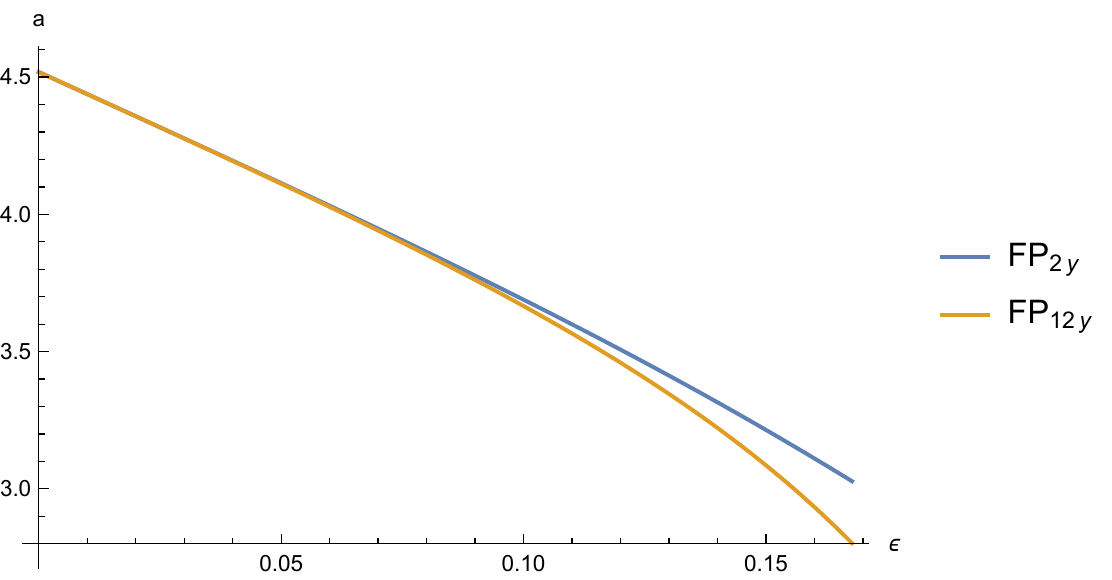} 
         \includegraphics[width=0.47\textwidth]{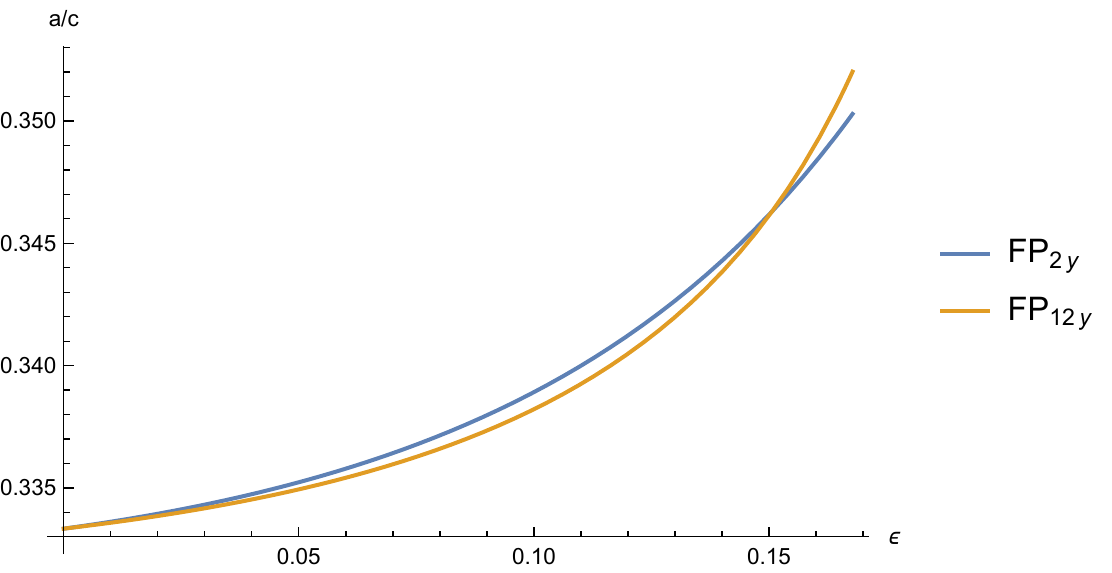}
    \caption{$a$ and $a/c$ as a functions of $\epsilon$ for the (semi)interacting fixed points.}
    \label{BLa}
\end{figure}

  With these charges we plot in Fig.~[\ref{BLa}] the value of $a$ and $a/c$ as functions of $\epsilon$. 

It is clear from the figure that all bounds are respected and that furthermore the highest value of $a$ is for  FP$_{2y}$ suggesting that if a flow exists between this and the fully interacting fixed point, it can be seen as an ultraviolet safe fixed point  along this trajectory. This is the susy equivalent of the phenomenon discovered in \cite{Esbensen:2015cjw}. In addition we also notice that the fully gaussian fixed point has the highest possible value of $a$ establishing an hierarchy of UV fixed points according to which, de facto, any phenomenological interesting field theory of this type would eventually flow to the fully gaussian one. This is substantially different from the case of \cite{Litim:2014uca} in which, at least perturbatively, the only UV fixed point  has the maximum $a$. In addition we expect no separatrix directly connecting FP$_{12y}$ with the gaussian fixed point  but a separatrix along the $\alpha_2$ coupling direction connecting it to FP$_{2y}$ because the linearised flow around the gaussian fixed point must necessarely coincide with the  perturbative analysis.

\section{Conclusions}
\label{conclusions}

We studied the short distance behaviour of several distinct classes of not asymptotically free supersymmetric gauge theories.   In particular we investigated super QCD with two adjoint fields and generalised superpotentials. Here we showed that nonperturbative asymptotic safety can be achieved without violating the known constraints provided the superpotentials assume specific forms. 

We also investigated the emergence of asymptotic safety within supersymmetric field theories featuring only gauge interactions. We discovered that asymptotic safety can be achieved at the cost of introducing a large enough number of matter fields in distinct representations of the gauge groups. In addition we investigated also semi-simple gauge theories with superpotentials such as quiver theories, and demonstrated that asymptotic safety can be achieved as well. Here the mechanism at play requires connecting the UV safe theory to an interacting IR one. 
 
  Our results integrate and extend  the initial work of Ref.~\cite{Intriligator:2015xxa} by introducing new mechanisms to achieve  supersymmetric safety.

\subsubsection*{Acknowledgments}
BB acknowledges the financial support from the Slovenian Research Agency (research core funding No.~P1-0035). The work of ND and FS is partially supported by the Danish National Research Foundation under the grant DNRF:90. BB thanks the CERN theoretical physics division for the hospitality.

\end{document}